\documentclass[11pt,a4paper]{amsart}
\pdfoutput=1
\usepackage[latin1]{inputenc}
\usepackage{hyperref}
\usepackage{amsmath,amsthm,cite,eucal,graphicx,subfig}
\usepackage[expansion=false]{microtype}

\newcommand{\myetal}{et al.\mbox{}}
\newcommand{\mysize}[1]{{\lvert #1 \rvert}}

\newcommand{\myC}{\mathcal{C}}
\newcommand{\myG}{\mathcal{G}}
\newcommand{\myH}{\mathcal{H}}
\newcommand{\myP}{\mathcal{P}}
\newcommand{\myfalse}{\text{false}}
\newcommand{\mytrue}{\text{true}}

\newtheorem{theorem}{Theorem}

\hypersetup{
    colorlinks=false,
    pdfborder={0 0 0},
    pdftitle={A simple local 3-approximation algorithm for vertex cover},
    pdfauthor={Valentin Polishchuk, Jukka Suomela},
}

\begin{document}

\title[]{A simple local 3-approximation algorithm \\ for vertex cover}
\author[]{Valentin Polishchuk}
\author[]{Jukka Suomela}
\address{Helsinki Institute for Information Technology HIIT \\
    Helsinki University of Technology and University of Helsinki}
\email{valentin.polishchuk@cs.helsinki.fi}
\email{jukka.suomela@cs.helsinki.fi}
\begin{abstract}
    We present a local algorithm (constant-time distributed algorithm) for finding a $3$-approximate vertex cover in bounded-degree graphs. The algorithm is deterministic, and no auxiliary information besides port numbering is required.
\end{abstract}
\maketitle

\section{Introduction}

Given a graph $\myG = (V,E)$, a subset of nodes $C \subseteq V$ is a \emph{vertex cover} if each edge $\{u,v\} \in E$ has $u \in C$ or $v \in C$. In this work, we present a constant-time distributed algorithm for finding a factor $3$ approximation of minimum vertex cover in bounded-degree graphs.

A distributed algorithm that runs in constant time (constant number of synchronous communication rounds) is called a \emph{local algorithm} \cite{naor95what}. In a local algorithm, the output of a node is a function of the input that is available within its constant-radius neighbourhood; this implies not only high scalability but also high fault-tolerance, making local algorithms desirable for real-world large-scale distributed systems.

Unfortunately, to date most results on local algorithms have been negative, even if we use Linial's \cite{linial92locality} model of distributed computing where the message size is unbounded and local computation is free. Linial's \cite{linial92locality} seminal work shows that there is no local algorithm for finding a maximal independent set, maximal matching, or 3-colouring of an $n$-cycle. This holds even if each node is assigned a unique identifier from the set $\{1,2,\dotsc,n\}$. Randomness does not help either; more generally, Naor and Stockmeyer \cite{naor95what} show that randomness does not help in so-called locally checkable labellings; maximal matching in a bounded-degree graph is an example of such a problem.

Kuhn \myetal{} \cite{kuhn05price,kuhn04what,kuhn06price} show that there is no local, constant-factor approximation algorithm for minimum vertex cover, minimum dominating set, or maximum matching in general graphs (without a degree bound). Randomness and unique node identifiers do not help.

Prior positive results on local algorithms for combinatorial problems typically rely on randomness, and the approximation guarantees only hold in expectation or with high probability. Wattenhofer and Wattenhofer \cite{wattenhofer04distributed} and Hoepman \myetal{} \cite{hoepman06efficient} present randomised local algorithms for weighted matching in trees; the algorithms provide a constant-factor approximation in expectation. Czygrinow \myetal{} \cite{czygrinow08fast} present a randomised local algorithm for finding a maximum independent set in a planar graph; the approximation ratio is $1+\epsilon$ with high probability. A more general framework for approximating covering and packing problems by local algorithms is based on solving the LP relaxation and applying randomised rounding \cite{kuhn05price,kuhn06price}.

Another line of research has studied local algorithms in a setting where auxiliary information is available. For example, if each node in a unit-disk graph knows its coordinates, then there is a local $(1+\epsilon)$-approximation of vertex cover \cite{wiese08local}.

However, without randomness or auxiliary information, positive results are scarce. Some deterministic local algorithms exist for linear programs \cite{floreen08tight,floreen08approximating,kuhn05price,kuhn06price}, but very few are known for combinatorial problems -- in the light of strong negative results, this is not particularly surprising. Naor and Stockmeyer \cite{naor95what} present a deterministic local algorithm for so-called weak colouring in graphs of odd degree. Lenzen \myetal{} \cite{lenzen08what} present a deterministic local $74$-approximation algorithm for minimum dominating set in planar graphs.

In this work, we give a new example of a simple, deterministic, constant-time, constant-factor approximation algorithm for a classical combinatorial problem; the algorithm does not resort to an LP approximation scheme and rounding. Our result is summarised in the following theorem.

\begin{theorem}\label{thm:main}
    A $3$-approximation to minimum vertex cover in a bounded-degree graph can be found by a deterministic local algorithm in $2 \Delta + 1$ communication rounds, where $\Delta$ is the maximum degree of the graph. The algorithm does not need unique node identifiers; port numbering is sufficient.
\end{theorem}

By \emph{port numbering} \cite{angluin80local} we mean that each node of $\myG$ imposes an ordering on its adjacent edges. Port numbering without any unique identifiers is an extremely weak assumption. For example, it does not help to break the symmetry in an $n$-cycle or an $n$-clique: in the worst case, every node is bound to make the same decision. In spite of that, we show that even in this very restricted model, it is possible to approximate the vertex cover to within a factor of $3$, which is not much worse than what can be obtained in a centralised setting by the best known polynomial-time approximation algorithms.

One explanation for this surprising positive result is the following. Indeed, we cannot break the symmetry in a symmetric graph. However, in a symmetric graph -- or, more generally, in a regular graph -- the trivial choice of all nodes is a factor $2$ approximation of vertex cover. Hence the instances that require a nontrivial choice are exactly those which cannot be entirely symmetric; there must be variation in the node degrees.

The only assumption that we make is some constant upper bound on the degree of the nodes. This is unavoidable, if we want a constant-time, constant-factor approximation algorithm for vertex cover \cite{kuhn05price,kuhn04what}.

\section{Overview}

To obtain a $2$-approximation of vertex cover in a centralised setting, one could simply find a maximal matching $M \subseteq E$ and output all matched nodes. Unfortunately, Linial's \cite{linial92locality} lower bound shows that the same technique cannot be applied in a local setting: even if unique node identifiers are available, we cannot find a maximal matching. However, Ha\'{n}\'{c}kowiak \myetal{} \cite{hanckowiak98distributed} show, in passing, that if the input graph is $2$-coloured (not only bipartite but also each node knows its part) then it is possible to overcome Linial's bound. Their distributed algorithm for maximal matching uses a subroutine called \emph{LowDegreeMatch}; this subroutine is a local algorithm for finding a maximal matching in bounded-degree $2$-coloured graphs.

\begin{figure}
    \subfloat[]{\label{fig:a}\includegraphics{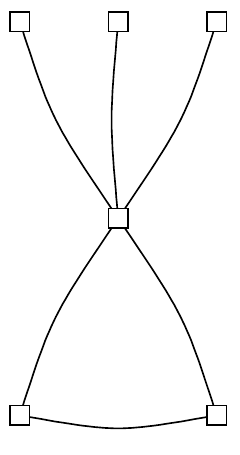}}%
    \hspace{\stretch{1}}%
    \subfloat[]{\label{fig:b}\includegraphics{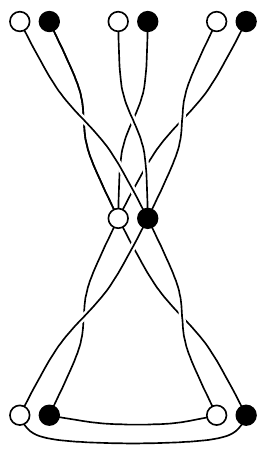}}%
    \hspace{\stretch{1}}%
    \subfloat[]{\label{fig:c}\includegraphics{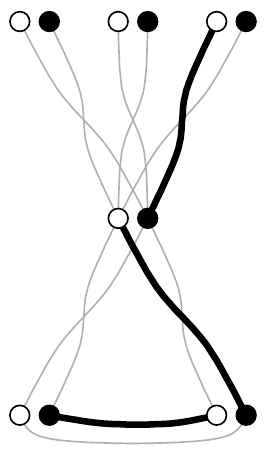}}%
    \hspace{\stretch{1}}%
    \subfloat[]{\label{fig:d}\includegraphics{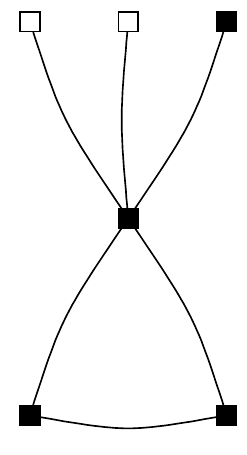}}
    \caption{Algorithm overview.}\label{fig:overview}
\end{figure}

How does this result help us though if we want to find a vertex cover in general (not 2-coloured) graphs? The idea is illustrated in Figure~\ref{fig:overview}. Given the graph $\myG$ (Figure~\ref{fig:a}), we replace each node with two copies, a black copy and a white copy. If the nodes $u$ and $v$ are adjacent in the original graph, then the black copy of $u$ is adjacent to the white copy of $v$ in the new graph, and vice versa. We obtain a bipartite, $2$-coloured graph $\myH$ (Figure~\ref{fig:b}). Now we can apply a local algorithm to find a maximal matching $M$ in the graph $\myH$ (Figure~\ref{fig:c}). Our approximate vertex cover for $\myG$ consists of those nodes whose either black copy or white copy (or both) were matched in $\myH$ (Figure~\ref{fig:d}). This turns out to be within factor $3$ of the optimum, because the edges of the matching in $\myH$ form a set of cycles and paths in~$\myG$.

We present the full algorithm in detail in Section~\ref{sec:alg}, and we prove the approximation guarantee in Section~\ref{sec:analysis}.

\section{Algorithm}\label{sec:alg}

We describe the local algorithm that finds a vertex cover $C \subseteq V$.

In the port numbering model, it is assumed that each node $v \in V$ knows its own degree $d(v) \le \Delta$. The node has $d(v)$ ports, each leading to one of its neighbours; the ports are numbered in an arbitrary order by $1, 2, \dotsc, d(v)$. A node can send a message to a given port, and the respective neighbour can receive it on the next time step.

The node $v \in V$ maintains the following variables: $a(v)$ and $b(v)$ are two chosen neighbours (identified by port numbers), and $i(v)$ is a counter. The output of the node is $c(v) \in \{ \mytrue, \myfalse \}$ which determines whether $v \in C$ or not.

Initially, $a(v) = \bot$, $b(v) = \bot$, $i(v) = 0$, and $c(v) = \myfalse$.\pagebreak

On an odd time step, each node $v \in V$ performs the following read--compute--write cycle.

\begin{enumerate}
    \item If $a(v) = \bot$ and $1 \le i(v) \le d(v)$, then receive a message $m$ from the port $i(v)$. If $m = \text{`accept'}$ then $a(v) \gets i(v)$ and $c(v) \gets \mytrue$.
    \item If $a(v) = \bot$ and $i(v) \le d(v)$ then $i(v) \gets i(v) + 1$.
    \item If $a(v) = \bot$ and $i(v) \le d(v)$ then send the message `propose' to the port $i(v)$.
\end{enumerate}

On an even time step, each node $v \in V$ performs the following read--compute--write cycle.

\begin{enumerate}
    \item Receive messages from all neighbours.
    \item For each $j$ such that a message `propose' was received from the port $j$, in increasing order:
    \begin{enumerate}
        \item If $b(v) = \bot$ then send the message `accept' to the port $j$. Set $b(v) \gets j$ and $c(v) \gets \mytrue$.
        \item Otherwise, send the message `reject' to the port $j$.
    \end{enumerate}
\end{enumerate}

Clearly, after $2\Delta+1$ time steps, the algorithm stops, as no messages are sent any more.

\section{Analysis}\label{sec:analysis}

Let us first show that the set $C = \{ v \in V : c(v) = \mytrue \}$ is a vertex cover when the algorithm stops. Consider an arbitrary edge $e = \{u, v\} \in E$. If $a(u) \ne \bot$, then $c(u) = \mytrue$. Otherwise $u$ has sent a `propose' message to $v$, and $v$ has sent a `reject' message; hence $b(v) \ne \bot$ and $c(v) = \mytrue$. We conclude that $C$ covers the edge $e$.

Let us now establish the approximation ratio. Let $C^{*}$ be a minimum vertex cover.

Let $v \in V$ be such that $a(v) \ne \bot$. Then the port $a(v)$ in $v$ leads to a node $u \in V$ such that $b(u) \ne \bot$. Furthermore, the port $b(u)$ in $u$ leads back to the node $v$. We say that $u$ and $v$ form a \emph{pair}.

Let $P \subseteq E$ consist of all edges $\{u,v\}$ such that $u$ and $v$ form a pair and consider the subgraph $\myG_1 = (V, P)$ of $\myG$. We make the following observations.
\begin{enumerate}
    \item The degree of a node $v \in V$ in $\myG_1$ is at most $2$. Indeed, at most one of its neighbours is determined by $a(v)$, and at most one of its neighbours is determined by $b(v)$.
    \item The set of non-isolated nodes (nodes with degree at least $1$) in $\myG_1$ is equal to the set $C$.
\end{enumerate}
Discard the isolated nodes to obtain the subgraph $\myG_2 = (C, P)$ of $\myG$. Each connected component of $\myG_2$ is a path or a cycle, and there are no isolated nodes.

Consider an arbitrary connected component $\myC$ of $\myG_2$. Either $\myC$ is a path $\myP$, or we can remove one edge arbitrarily to obtain a path. The paths form a partition of the cover $C$; each $v \in C$ belongs to exactly one such path.

Let $m \ge 1$ be the number of edges on the path $\myP$. As $\myP$ is a subgraph of $\myG$, each edge of $\myP$ must have at least one endpoint in the optimal cover $C^{*}$. Hence at least $\lceil m/2 \rceil$ nodes of $\myP$ are in $C^{*}$, which is at least a fraction $1/3$ of the total number of nodes in $\myP$ (the worst case being $m = 2$).

Summing over all paths, we conclude that $\mysize{C} \le 3 \mysize{C^{*}}$. This completes the proof of Theorem~\ref{thm:main}.

\section{Discussion}

We presented a simple local algorithm for finding a factor $3$ approximation for vertex cover in bounded-degree graphs. The algorithm does not use LP rounding, and can be easily implemented. One challenge for future work is closing the gap between the running time $O(\Delta)$ of our algorithm and the lower bound $\Omega(\log \Delta / \log \log \Delta)$ from prior work \cite{kuhn05price,kuhn04what}.

\section*{Acknowledgements}

We thank Patrik Floréen and Petteri Kaski for discussions and comments. This research was supported in part by the Academy of Finland, Grants 116547 and 118653 (ALGODAN), and by Helsinki Graduate School in Computer Science and Engineering (Hecse).

\providecommand{\noopsort}[1]{}

\end{document}